\documentclass{article} 
\usepackage{iclr2027_conference,times}

\usepackage{amsmath}
\usepackage{graphicx}
\usepackage{hyperref}
\usepackage{url}
\usepackage{booktabs}
\usepackage{tabularx}
\usepackage{enumitem}
\usepackage{multirow}
\usepackage{adjustbox}
\usepackage{listings}
\lstdefinestyle{promptbox}{
  basicstyle=\footnotesize\ttfamily,
  breaklines=true,
  breakatwhitespace=false,
  columns=fullflexible,
  keepspaces=true,
  frame=single,
  framesep=6pt,
  xleftmargin=8pt,
  xrightmargin=8pt,
  aboveskip=4pt,
  belowskip=4pt
}

\iclrfinalcopy

\begin{document}
\title{Benchmarking Coding Agents for System-Level Performance Optimization}

\author{%
    \textbf{Yanze Zhang} \& \textbf{Maggie Yongqi Guan} \& \textbf{Shudong Liu} \& \textbf{Qi Hao} \\
  \textbf{Kanye Ye Wang} \& \textbf{Meng Qu} \& \textbf{Xiaobo Zhou} \\
  Faculty of Information Science and Computing \\
  University of Macau \\
  Macau SAR, China \\
  \texttt{\{yc47943,yc37963,yc27405,qihao,wangye,mengqu,waynexzhou\}@um.edu.mo} \\
}

\maketitle

\lhead{\small Preprint. Under review as a conference paper at ICLR 2027}

\begin{abstract}

Coding agents are becoming an increasingly important paradigm for automating software engineering tasks. However, existing coding benchmarks primarily evaluate agents at the function or repository level, leaving their capabilities in more complex distributed software systems largely unexplored. In practice, many modern applications adopt microservice architectures, where end-to-end performance emerges from interactions among multiple independently running services, requiring agents to reason beyond local code changes.
To study this capability, we introduce MicroPerf-Bench, a benchmark for evaluating coding agents on end-to-end performance optimization in microservice systems. Constructing such a benchmark is challenging because reproducible system-level performance fixes are sparse and difficult to attribute. We address this challenge by injecting empirically grounded performance anti-patterns into real-world open-source systems while preserving functionality and ensuring measurable performance degradation. The resulting benchmark contains 65 optimization instances across seven systems and nine anti-patterns, each with an executable environment, functional tests, and mixed-workload evaluation.
Across 1,755 runs covering nine agent–model configurations, the best configuration recovers only 42.0\% of the injected performance degradation. The dominant failure is not invalid code, but functionally correct and constraint-compliant changes that fail to improve end-to-end performance, highlighting system-level bottleneck localization as a key challenge for current coding agents. The source code and data are available at \url{https://github.com/zyzcjnb/microperf}.

\end{abstract}


\section{Introduction}

Large language models have rapidly evolved from code-generation tools into coding agents that can solve software engineering tasks~\citep{jimenez2024swebench,yang2024sweagent}. As these agents become increasingly capable, reliably evaluating their capabilities becomes increasingly important. Existing coding-agent benchmarks have largely focused on \emph{functional} software engineering tasks. For example, SWE-bench~\citep{jimenez2024swebench} evaluates whether agents can resolve software issues by producing patches that satisfy functional tests. Such benchmarks are crucial for measuring progress in coding agents, but functional correctness captures only one dimension of software quality.

Recently, there has been growing interest in a complementary capability: \emph{performance optimization}, which asks whether an agent can make software more efficient in terms of execution time and resource usage while preserving its functionality. Such software performance depends on the workload, execution environment, and runtime interactions. Therefore, an effective optimization agent should identify the source of performance degradation, determine an appropriate modification, and verify that the modification produces a measurable improvement under execution.

Motivated by this challenge, recent work has begun to benchmark the performance-optimization capabilities of coding agents. Early benchmarks mainly study performance at the function or program level. For example, EffiBench~\citep{huang2024effibench} evaluates the efficiency of generated implementations on self-contained programming tasks. More recent benchmarks, such as GSO~\citep{shetty2026gso}, extend evaluation to the repository level, requiring agents to navigate larger codebases and produce correctness-preserving optimizations. However, these settings still largely treat a program or codebase together with its workload as the unit of optimization. An important capability remains less explored: \emph{system-level performance optimization}, where end-to-end behavior emerges from interactions among multiple components during execution.

Microservice systems provide a representative setting for studying this capability. A microservice application is composed of multiple independently running services that communicate through network calls. As a result, the latency experienced by a user may depend on an entire execution path across services. Performance optimization in this setting thus requires agents to reason beyond local code changes and understand how components interact at runtime. However, building a benchmark in the microservice setting is nontrivial. Reproducible real-world system-level performance fixes are sparse, often intertwined with other code changes and difficult to attribute. At the same time, arbitrarily synthesized performance bugs may fail to reflect the problems that arise in real systems.

To address these challenges, we introduce \textbf{MicroPerf-Bench}, a benchmark for evaluating coding agents on end-to-end performance optimization in microservice systems. Rather than relying solely on mined fixes, we construct benchmark instances through controlled injection of empirically grounded microservice performance anti-patterns~\citep{keck2016antipattern}, which capture recurring system-level inefficiencies. We inject them into real-world microservice systems while preserving functional behavior and ensuring measurable performance degradation. The resulting benchmark contains 65 optimization instances across seven microservice systems and nine performance anti-patterns. Each instance provides an executable multi-service environment, functional tests, and mixed-workload performance evaluation. Given an under-optimized system, an agent must inspect and modify its implementation. The affected services are then rebuilt and redeployed, and the resulting system is evaluated for both functional correctness and end-to-end performance improvement.

Using MicroPerf-Bench, we conduct a systematic evaluation of nine agent--model configurations formed by three agent harnesses and three underlying models. We run every configuration three times on all 65 instances, yielding 1,755 optimization runs. Current coding agents remain far from solving the benchmark: even the best configuration achieves an average normalized score of only 42.0 out of 100. More importantly, our failure analysis reveals that the dominant failure is not invalid code. Instead, agents frequently produce functionally correct modifications that yield no measurable end-to-end performance gain. This finding exposes an important gap between agents' ability to \emph{modify code} and their ability to \emph{understand and optimize system-level runtime behavior}: identifying the true bottleneck remains a central challenge. Our broader analysis further shows that performance varies substantially across host systems and repeated runs, motivating both system-aware optimization methods and more robust evaluation protocols.

\section{Background and Motivation}
\label{sec:background}

\textbf{Performance in Microservice Systems.}
Microservice architectures have become the dominant paradigm for modern software systems, enabling developers to build, scale, and deploy complex applications as collections of small, loosely coupled, and independently deployable services~\citep{newman2021microservices,gan2019open}. Despite their operational benefits, microservice architectures fundamentally alter the nature of software performance optimization.

In traditional monolithic software, performance bottlenecks are typically localized to specific algorithmic inefficiencies, unoptimized database queries, or excessive I/O within a single process. In contrast, performance in a microservice system is an \emph{emergent property} arising from the dynamic interactions among multiple services executing across a distributed network. A single user-facing request often triggers a directed acyclic graph (DAG) of downstream inter-service remote procedure calls (RPCs), asynchronous message exchanges, and shared database transactions. Consequently, end-to-end user latency is determined not merely by individual code efficiency, but by cross-service communication overhead, service dependency topologies, serialization bottlenecks, and resource contention. Optimizing a microservice system thus requires agents to reason beyond localized functions and comprehend system-wide runtime execution paths.

\textbf{Limitations of Existing Coding Benchmarks.}
Recent advances in large language models (LLMs) have driven the rapid development of autonomous coding agents capable of resolving real-world software engineering issues~\citep{jimenez2024swebench,yang2024sweagent}. To measure progress, early performance optimization benchmarks evaluated models on standalone function-level tasks~\citep{huang2024effibench}, while recent benchmarks such as GSO~\citep{shetty2026gso} and SWE-perf~\citep{he2025swe} extended evaluations to repository-level codebases. However, existing benchmarks operate under the assumption that a program can be compiled, tested, and profiled as a self-contained unit within a single codebase. This setup fails to model microservice environments due to two gaps:

\emph{(1) Cross-Service Structural Reasoning Gap:} In a distributed architecture, fixing an end-to-end bottleneck often requires structural modifications spanning multiple repositories or services—such as introducing an API gateway, replacing synchronous RPCs with asynchronous queues, or consolidating chatty interfaces. Repository-level benchmarks cannot formulate or evaluate such cross-boundary architectural refactoring.

\emph{(2) Deployment-Centric Evaluation Gap:} Local code efficiency does not guarantee system-level performance gains. A change that speeds up a local function might simply shift the system bottleneck to a downstream database or alter concurrency dynamics under heavy traffic. Accurately assessing microservice performance requires deploying the complete multi-service stack (e.g., via container orchestration) and subjecting it to dynamic, end-to-end multi-endpoint workloads.

\textbf{Performance Anti-Patterns as Ground Truth.}
To bridge this gap and establish a rigorous benchmark, we draw upon the concept of \emph{performance anti-patterns}~\citep{keck2016antipattern, cerny2023catalog}. Performance anti-patterns represent recurring architectural and implementation bad practices that consistently induce severe system-level performance degradation under load. Examples include \emph{Chatty Service} (excessive fine-grained RPC calls) and \emph{Shared Persistency} (multiple services coupling directly to a single database). 

Utilizing empirically grounded performance anti-patterns offers an ideal framework for benchmark construction: they provide realistic, real-world-relevant performance bugs, have verifiable optimization paths through code and topology refactoring, and enable precise attribution of performance recovery under controlled experimental conditions.


\section{MicroPerf-Bench}
\label{sec:microperf_bench}


To systematically evaluate coding agents on end-to-end performance optimization in distributed systems, we present \textbf{MicroPerf-Bench}. The benchmark comprises 65 system-level optimization instances constructed from seven real-world open-source microservice applications and nine performance anti-patterns. Each instance provides a fully deployable multi-container environment, functional test suites, and end-to-end workload profiling scripts. 

Next, we formalize the optimization task (Section~\ref{sec:task_formulation}), describe our rigorous data construction and validation methodology (Section~\ref{sec:data_construction}), and detail the evaluation protocols and metrics (Section~\ref{sec:bench_eval}).

\subsection{Task Formulation}
\label{sec:task_formulation}





We formulate system-level performance optimization as a constrained patch-generation task. 

\textbf{Input.} For each benchmark instance $i$, an agent is provided with: (1) The complete under-optimized multi-service codebase $\mathcal{C}_i$; (2) Container deployment configurations $\mathcal{M}_i$ (e.g., Docker Compose manifests); (3) An executable mixed-workload load testing script $\mathcal{W}_i$ reproducing user traffic; (4) A formal task specification $\mathcal{S}_i$ outlining the target application and system boundary constraints (Appendix~\ref{app:taskspec}).

\textbf{Constraints \& Anti-Shortcut Rules.} To ensure that improvements originate from genuine code or architectural optimizations rather than artificial shortcuts, $\mathcal{S}_i$ explicitly prohibits agents from: (1) increasing hardware resource allocations (e.g., boosting CPU/memory limits in Docker); (2) disabling non-functional system mechanisms (e.g., dropping distributed tracing, logging, rate-limiting, or authentication middleware); or (3) weakening database transaction safety or security validation.

\textbf{Output.} The agent must generate a set of modified or newly added source code and configuration files $\Delta \mathcal{C}_i$. The benchmark interface accepts standard unified file diffs or directory artifacts, ensuring complete compatibility with arbitrary agent scaffolds and LLMs.

\textbf{Deployment-Centric Evaluation Execution.} Upon receiving $\Delta \mathcal{C}_i$, the benchmark execution harness rebuilds the affected container images, redeploys the distributed system in an isolated environment, and executes both functional correctness tests and dynamic workload profiling.

\begin{figure}[t]
  \centering
  \includegraphics[width=0.8\linewidth]{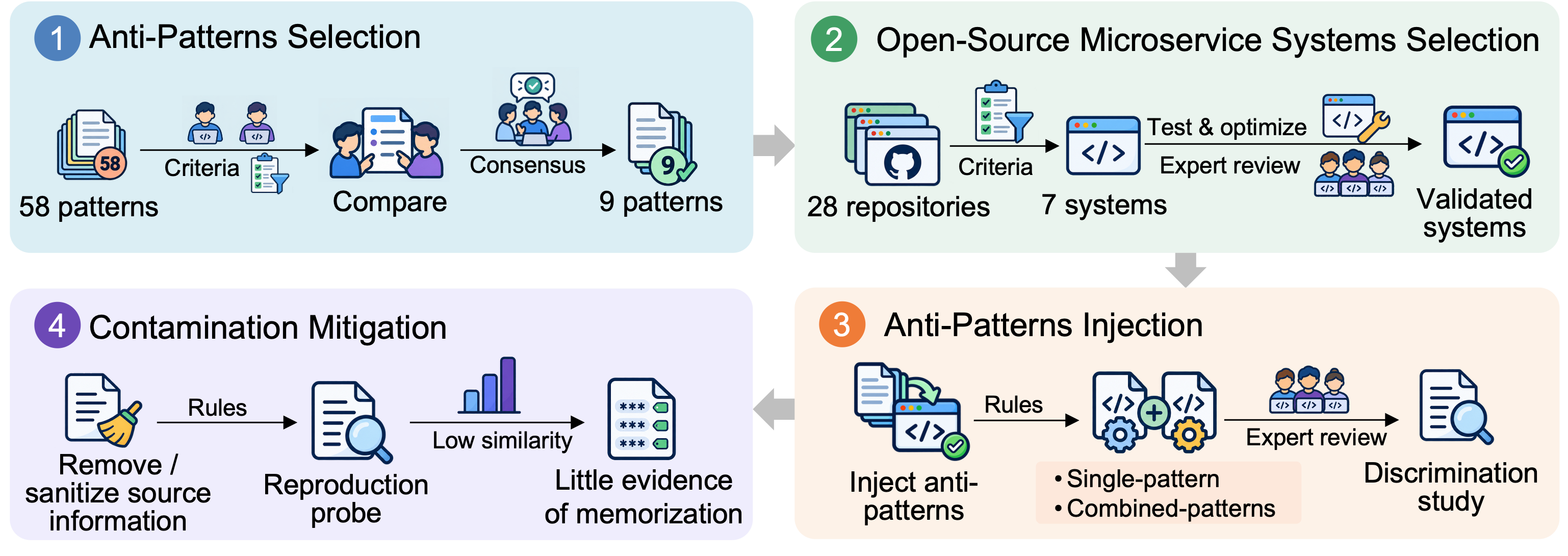}
  \caption{The four-phase data construction pipeline of MicroPerf-Bench.}
  \label{fig:pipeline}
\end{figure}


\subsection{Data Construction}
\label{sec:data_construction}

Constructing a reliable system-level performance benchmark presents a key dilemma: mined real-world commits are sparse and entangled, while synthetic bugs risk being unrealistic or trivial. 

In a pre-study, we audited over 10,000 commits across 12 popular open-source microservice repositories (Appendix~\ref{app:mapping}). Only 15 commits give pure, attributable microservice performance fixes, with 9 concentrated in a single suite with specialized maintenance. Moreover, real fixes often bundle feature additions or framework upgrades, making isolated performance attribution nearly impossible. 

To solve this, we adopt a controlled injection methodology anchored in empirically grounded performance anti-patterns. Figure~\ref{fig:pipeline} details our four-phase construction pipeline.

\textbf{Phase I: Performance Anti-Pattern Selection \& Real-World Grounding.}
Starting from 58 catalogued microservice anti-patterns and bad smells~\citep{cerny2023catalog}, we applied three rigorous screening criteria. (1) \emph{Performance Relevance}: The pattern must produce measurable runtime latency degradation rather than mere code maintainability issues. (2) \emph{Reproducible Degradation}: The latency impact must significantly exceed environmental measurement noise under standard workloads. (3) \emph{Optimization Actionability}: The pattern must be fixable via code, architectural, or topology refactoring (excluding hardware scaling).

This process yielded nine target anti-patterns: \emph{Nano-service}, \emph{Mega service}, \emph{Service chain}, \emph{Chatty service}, \emph{Shared persistency}, \emph{Empty messages}, \emph{On-line only}, \emph{Hub-like dependency}, and \emph{ESB misuse} (retained in our taxonomy as an uninstantiated coverage gap). All selected patterns directly mirror documented industrial challenges (Appendix~\ref{app:mapping}). For example, \emph{Chatty Service} drove the adoption of GraphQL and Backend-for-Frontend (BFF) patterns at Meta and SoundCloud~\citep{graphql2015, soundcloud-bff}; \emph{Shared Persistency} mirrors database contention reported by Discord~\citep{discord2023}; \emph{On-line Only} motivated Netflix's transition to asynchronous cloud gateways~\citep{zuul2016}.

\textbf{Phase II: Target System Selection \& Baseline Cleaning.}
We evaluated 28 open-source microservice repositories against four criteria: active maintenance, thorough API documentation, reproducible mixed-workload scripts, and architectural diversity. Seven representative applications were selected: Spring PetClinic Microservices, MartianBank Demo, Robot Shop, Pitstop, and three DeathStarBench suites (Social Network, Media Service, Hotel Reservation).

To ensure that evaluation scores strictly measure the recovery of \emph{injected} degradation, we established clean "Original Reference" ($R$) baselines free of native defects. Three senior software engineers with over three years of microservice industry experience independently audited the systems, resolving pre-existing issues such as N+1 query patterns and missing database indexes. Fleiss' $\kappa$ inter-annotator agreement reached $0.87$ ($92.3\%$ agreement), resolving 15 native defects. Additionally, fixed client-side ``think times'' were removed from workload scripts to ensure server-side latency bottlenecks were transparently exposed.

\textbf{Phase III: Controlled Anti-Pattern Injection \& Quality Validation.}
We injected anti-patterns under three strict constraints: (i) preserve all externally exposed API contracts; (ii) maintain full functional completeness; and (iii) restrict modifications to service decomposition, instance topology, communication protocols, messaging patterns, and data-access layers. 

Injections proceeded in two stages. (1) \emph{Single-Pattern Injections}: Each single-pattern variant was required to exhibit at least a 20\% end-to-end latency degradation relative to $R$ to guarantee a strong signal-to-noise ratio. (2) \emph{Combined-Pattern Injections}: Validated single patterns were co-injected into multi-service systems to simulate complex real-world multi-bottleneck pathologies.

After injection, all 65 instances were independently validated by three experts with industrial experience in microservice development.
Reviewers successfully diagnosed the injected anti-pattern from runtime behavior alone in $81.5\%$ of cases.
We further conducted a blinded discrimination study with 12 injected instances and 12 real parent-commit defects, where the accuracy in identifying the defect source was $47.2\%$, suggesting that the injected instances did not exhibit obvious artificial cues under this evaluation setting.
Details are provided in Appendix~\ref{app:expert_validation}.

\textbf{Phase IV: Contamination Mitigation.}
\label{pha:contaminationmitigation}
Because the base open-source repositories exist in public datasets, agents might theoretically improve scores by memorizing upstream code. We sanitized all README files, code documentation, URLs, class identifiers, comments, and Git commit histories. Verbatim reproduction probing across evaluated LLMs revealed token-level overlap of only $7$--$10\%$ (worst-case $<25\%$), confirming that memorization cannot solve the benchmark.

\subsection{Benchmark Evaluation Protocol and Metrics}
\label{sec:bench_eval}

Evaluations are conducted in heavily isolated environments. Between runs, we batch-build images, purge environmental state, and launch fresh containers. Performance tests incorporate a 90-second warm-up phase to absorb JVM/container startup effects, followed by three averaged measurements.

\textbf{Isolated Evaluation Environment.}
Each evaluation run executes in a fully isolated container environment. Prior to profiling, the harness purges cached states, batch-builds fresh container images, and launches the services. To eliminate cold-start artifacts and JVM JIT compilation noise, each test runs a 90-second warm-up phase, followed by three averaged measurement iterations.

\textbf{Metric Selection Rationale.}
Throughput metrics (RPS) were found to plateau under non-saturated workloads and exhibit high hardware-dependent variance under saturated conditions. Consequently, we employ \emph{average end-to-end request latency under mixed workloads} as our primary performance metric. The workload exercises multiple API endpoints simultaneously, preventing agents from trivially overfitting to a single endpoint.

\textbf{Correctness and Fairness Gates.}
An agent's patch $O$ must pass two non-negotiable gates:
\begin{enumerate}[leftmargin=*, nosep]
    \item \emph{Functional Correctness Gate:} The optimized system must pass $100\%$ of functional API test suites and data-consistency assertions. Additionally, the request error rate under load must not exceed the degraded version's error rate by more than $1\%$.
    \item \emph{Fairness Compliance Gate:} Patches that achieve speedups by disabling security checks, rate limiting, distributed tracing, or transaction isolation are flagged as violations and assigned zero.
\end{enumerate}

\textbf{Core Scoring Mechanism.}
For each instance $i$, let $L_D$ represent the average end-to-end latency of the Degraded Baseline ($D$), $L_R$ represent the latency of the Original Reference ($R$), and $L_O$ represent the latency of the agent's Optimized Version ($O$). The normalized recovery score $S_i$ measures the percentage of injected degradation recovered by an agent and is calculated as $S_i = \mathrm{clamp}\left( 100 \times \frac{L_D - L_O}{L_D - L_R},\ 0,\ 100 \right)$, where $\mathrm{clamp}(x, 0, 100) = \max(0, \min(100, x))$. A score of $0$ indicates no improvement over the degraded baseline (or a gate failure), while a score of $100$ means full recovery to the clean reference performance. The overall benchmark performance $S_{\text{overall}}$ is the unweighted arithmetic mean across all instances, given by $S_{\text{overall}} = \frac{1}{N} \sum_{i=1}^{N} S_i$.


\section{Experiments}
\label{sec:exp}

To assess the capabilities of current coding agents in optimizing microservice performance, we conduct an extensive evaluation across multiple models and agent scaffolds. This section details our experimental setup, analyzes the overall performance and failure modes of baseline configurations, investigates the impact of runtime profiling, and provides qualitative case studies to illustrate agent behaviors.

\subsection{Evaluation Setup}
\label{sec:exp:setup}

\textbf{Agent Configurations.}
We evaluate nine baseline configurations formed by the Cartesian product of three agent scaffolds (OpenCode, OpenHands, and Claude Code) and three state-of-the-art base models (GLM-5.2, Qwen3.7-Max, and DeepSeek-V4-Pro). To encourage exploration and reasoning, all models operate with a temperature of $T{=}1.0$ and have their respective ``thinking'' or chain-of-thought modes enabled. Complete configuration details are provided in Appendix~\ref{app:agent_config}.

\textbf{Execution Protocol.}
Each configuration is provided with an identical task specification, which explicitly outlines the optimization objective, the permitted scope of modifications, and the correctness and fairness gates defined in Section~\ref{sec:bench_eval}. Agents can execute the provided mixed-workload script during optimization, but the task specification does not explicitly instruct them to perform performance measurements.
We allocate a total time budget of 2 hours per benchmark instance, with a strict 20-minute timeout per executed command. Because agentic optimization is inherently stochastic and the resulting score distribution is zero-inflated, we independently run each of the nine configurations three times across all 65 benchmark instances. This yields a total of $1{,}755$ independent optimization trials for our primary evaluation.


\textbf{Evaluation Metrics.}
We quantify optimization success using three complementary metrics, all averaged over the three independent runs unless otherwise specified:
\begin{itemize}[leftmargin=*, nosep]
    \item \emph{Average Score:} The mean of the instance-level scores ($S_i \in [0, 100]$) computed via Eq.~1. Trials that violate functional correctness or fairness constraints automatically receive a score of zero.
    \item \emph{Positive-Score Rate:} The percentage of trials yielding $S_i > 0$, indicating that the agent achieved a measurable latency reduction relative to the degraded baseline while satisfying all constraints.
    \item \emph{Full-Score / Partial-Score Rates:} We decompose the positive-score rate into the \emph{full-score rate} ($S_i = 100$, indicating complete performance recovery) and the \emph{partial-score rate} ($0 < S_i < 100$).
\end{itemize}


\textbf{Measurement Environment.}
All experiments are executed on a dedicated dual-socket server equipped with $2\times$ Intel Xeon Platinum 8562Y+ CPUs and 384\,GB of RAM, running Ubuntu 22.04. To ensure pristine measurements, every performance evaluation rebuilds the instance, deploys it in fresh containers, and executes a 90-second warm-up phase to stabilize JVM and container states. End-to-end latency is measured three times under the mixed workload, and the results are averaged.

\subsection{Overall Baseline Performance}
\label{sec:exp:overall}

Table~\ref{tab:overall} presents the performance of the baseline configurations across the trials. The results show that end-to-end system performance optimization remains a formidable challenge for coding agents. 

Average scores range from a low of 23.4 (Claude Code + DeepSeek) to a high of 42.0 (OpenHands + GLM). Across all configurations, the mean positive-score rate sits at just 44.0\%. More strikingly, full-score rates remain exceptionally low, ranging from 9.2\% to 22.6\%. This indicates that more than half of all optimization attempts fail to yield any valid performance improvement. 

Furthermore, when agents do succeed in improving performance, they frequently capture only a fraction of the potential gains. Partial-score outcomes account for 27.4\% of all trials, substantially outnumbering the 16.7\% of trials that achieve a full score. This discrepancy suggests that while agents can sometimes identify and mitigate secondary inefficiencies, they often miss the primary system bottleneck responsible for the bulk of the injected degradation. Finally, these baseline results expose a significant gap between agents' code-generation capabilities and their ability to optimize distributed systems, motivating the exploration of runtime feedback mechanisms (see Section~\ref{sec:exp:profiling}).

\begin{table*}[t]
\centering
\small
\caption{Overall performance of the nine configurations (three-run means over 65
instances). \emph{Avg.\ Score} is the mean percentage of injected degradation
recovered (Eq.~1); \emph{SD} is the standard deviation of the three per-run scores
(listed under \emph{Per-run}). The positive-score rate is decomposed into \emph{Partial}
($0 < S < 100$) and \emph{Full Score} ($S = 100$, performance fully restored to the
reference level). $\dagger$: single run (Section~\ref{sec:exp:profiling}).}
\label{tab:overall}
\setlength{\tabcolsep}{5pt}
\renewcommand{\arraystretch}{0.9}
\begin{adjustbox}{max width=\textwidth}
\begin{tabular}{llccccc}
\toprule
\multirow{2.5}{*}{\textbf{Scaffold}}
& \multirow{2.5}{*}{\textbf{Model}}
& \multirow{2.5}{*}{\textbf{Avg.\ Score}}
& \multirow{2.5}{*}{\textbf{SD}}
& \multirow{2.5}{*}{\textbf{Per-run}}
& \multicolumn{2}{c}{\textbf{Positive-Score Rate (\%)}} \\
\cmidrule(lr){6-7}
& & & & & \textbf{Partial} & \textbf{Full Score} \\
\midrule
\multirow{3}{*}{OpenCode}
 & GLM      & 40.2 & 5.9 & 46.6\,/\,34.8\,/\,39.3 & 33.4 & 17.4 \\
 & Qwen     & 31.8 & 5.0 & 32.9\,/\,26.3\,/\,36.2 & 24.6 & 14.9 \\
 & DeepSeek & 29.4 & 2.5 & 29.2\,/\,32.1\,/\,27.0 & 24.1 & 12.8 \\
\addlinespace[3pt]
\multirow{3}{*}{OpenHands}
 & GLM      & \underline{42.0} & 5.3 & 46.0\,/\,44.1\,/\,36.0 & 31.8 & 21.5 \\
 & Qwen     & 39.2 & 1.5 & 39.5\,/\,37.6\,/\,40.5 & 33.8 & 19.5 \\
 & DeepSeek & 29.8 & 1.6 & 31.6\,/\,29.3\,/\,28.4 & 22.5 & 15.4 \\
\addlinespace[3pt]
\multirow{3}{*}{Claude Code}
 & GLM      & 38.3 & 1.0 & 39.3\,/\,38.2\,/\,37.4 & 30.3 & 16.9 \\
 & Qwen     & 39.4 & 1.3 & 40.1\,/\,40.2\,/\,37.9 & 23.0 & \underline{22.6} \\
 & DeepSeek & 23.4 & 3.7 & 19.1\,/\,25.2\,/\,25.8 & 22.6 & 9.2 \\
\midrule
\multicolumn{2}{l}{Overall Mean} & 34.8 & \textemdash & \textemdash & 27.4 & 16.7 \\
\midrule
Claude Code & Qwen + profiling$^{\dagger}$ & \textbf{47.7} & \textemdash & 47.7 & 27.7 & \textbf{30.8} \\
\bottomrule
\end{tabular}
\end{adjustbox}
\end{table*}



\subsection{Effects of Base Model and Agent Scaffold}
\label{sec:exp:factors}

To isolate the contributions of underlying base models and agent scaffolds, we run paired $t$-tests on the per-instance mean scores across all 65 instances, using a significance threshold of $\alpha = 0.05$.

\textbf{Base Model Impact.} 
The choice of base model significantly influences optimization success. Averaged across all scaffolds, GLM outperforms DeepSeek by a substantial margin of 12.7 points ($p = 0.001$, Cohen's $d_z = 0.45$), while Qwen surpasses DeepSeek by 9.3 points ($p = 0.003$, $d_z = 0.39$). GLM also achieves a 3.4-point advantage over Qwen, though this difference falls short of statistical significance ($p = 0.105$). Notably, DeepSeek consistently ranks lowest across all three evaluated scaffolds, demonstrating a robust performance hierarchy among the models.

\textbf{Agent Scaffold Impact.} 
In contrast, performance variance acrossscaffolds is much smaller and less definitive. Averaged across models, OpenHands edges out OpenCode by 3.2 points ($p = 0.139$) and Claude Code by 3.3 points ($p = 0.133$). OpenCode and Claude Code perform almost identically, with a nominal 0.1-point difference ($p = 0.957$). None of these scaffold comparisons reach the significance threshold. Furthermore, the optimal scaffold appears model-dependent: OpenHands pairs best with GLM and DeepSeek, whereas Claude Code yields the highest score when paired with Qwen. While these results do not prove equivalence, they suggest that in the context of system-level performance tuning, the reasoning capacity of the base model outweighs the scaffolding logic.

\subsection{Failure Mode Analysis}
\label{sec:exp:failure}

To systematically understand why agents fail to optimize these systems, we classify all 982 zero-score trials into five distinct failure modes based on comprehensive audit logs: \textbf{no performance gain} (changes are correct, but fail to reduce latency), \textbf{fairness violation}, \textbf{functional correctness error}, \textbf{build or startup failure}, and \textbf{timeout}. Precise definitions are provided in Appendix~\ref{app:failure_detail}.

\begin{figure*}[t]
\centering
\includegraphics[width=\textwidth]{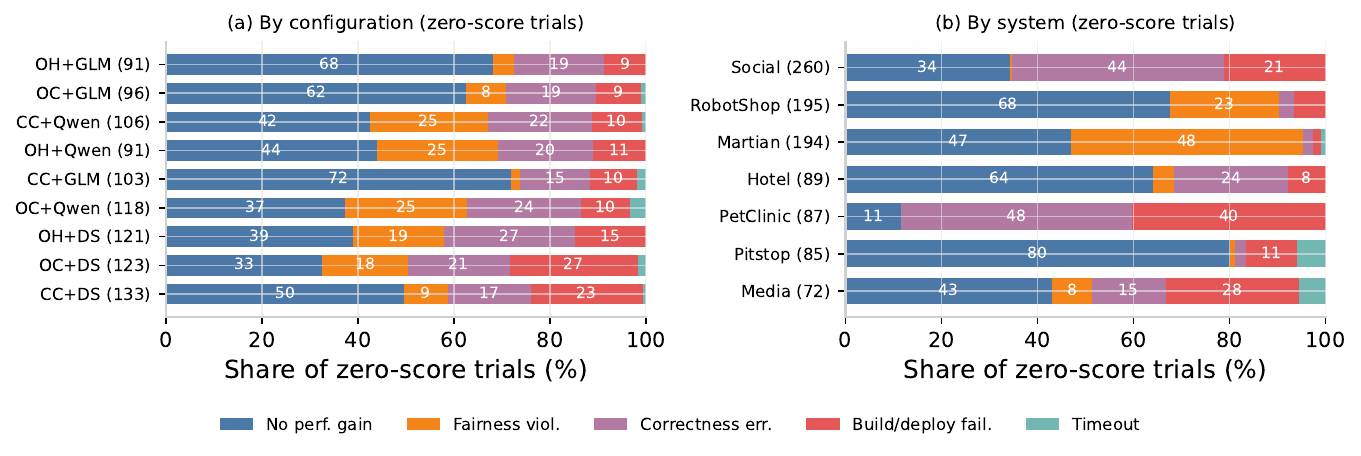}
\caption{
Failure-mode composition among zero-score trials.
(a) Results by configuration, ordered by average score.
(b) Results by system, ordered by the number of zero-score trials.
Counts in parentheses indicate the number of zero-score trials in each group.
Percentages are computed within each group's zero-score trials.}
\label{fig:failure}
\end{figure*}

As depicted in Figure~\ref{fig:failure}, the ``no performance gain'' mode dominates, accounting for 48.7\% of all failures. This is a critical insight, and agents frequently succeed at writing functionally correct, compilable code that adheres to all constraints, yet they fundamentally fail to move the needle on end-to-end latency. This suggests that, beyond code generation, \emph{bottleneck identification}, i.e., the ability to diagnose where time is actually being spent, remains a key challenge for current agents.

The distribution of failure modes is heavily influenced by the model. The ``no gain'' outcome constitutes 67.6\% of GLM's failures, indicating it safely edits code but often targets the wrong components. Conversely, Qwen and DeepSeek are more prone to fairness violations (e.g., stripping out vital tracing or validation logic to artificially speed up execution), with DeepSeek also suffering from higher rates of build failures. We also observe system-dependent failure profiles: instances in RobotShop and Pitstop fail almost exclusively due to ``no gain'' (67.7\% and 80.0\%, respectively), whereas PetClinic and Social Network trigger more functional correctness errors.

\subsection{Profiling-Tool Augmentation}
\label{sec:exp:profiling}

Motivated by the prevalence of ``no gain'' failures, we investigate if equipping agents with dynamic runtime feedback can improve their optimization efficacy. We integrate a dedicated profiling tool that supports automated workload testing, performance profiling, and distributed trace extraction.

\begin{figure*}[htbp]
\centering
\includegraphics[width=\textwidth]{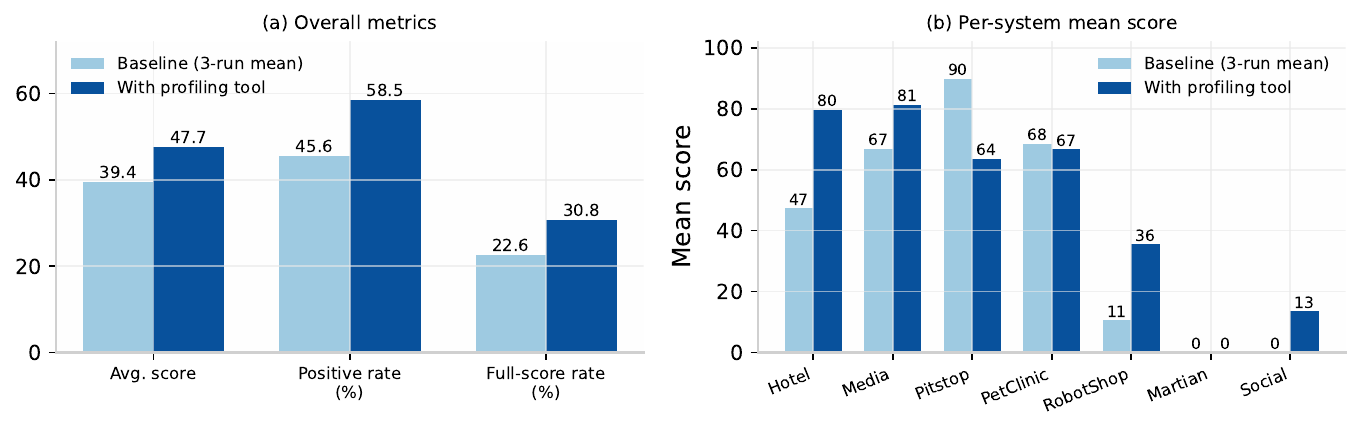}
\caption{Effect of profiling-tool augmentation on Claude Code + Qwen.
(a) Overall metrics for the three-run baseline mean and the single augmented run.
(b) Per-system score changes relative to the baseline mean.
Improvements vary across systems, with no change on Martian and decreases on PetClinic and Pitstop.
}
\label{fig:profiling}
\end{figure*}

We evaluate the Claude Code + Qwen configuration augmented with this tool in a single run across all 65 instances. The preliminary results are highly promising. As shown in Figure~\ref{fig:profiling}(a) and Table~\ref{tab:overall}, the augmented run achieves an average score of 47.7, comfortably surpassing its baseline mean of 39.4 and beating the highest baseline configuration across all nine setups (42.0). Its positive-score rate surges from 45.6\% to 58.5\%, and its full-score rate jumps from 22.6\% to 30.8\%.

However, the benefits of profiling are not uniform. Figure~\ref{fig:profiling}(b) reveals that while gains are substantial on systems like Hotel and RobotShop, performance on Pitstop actually regresses. This suggests that while dynamic feedback can rapidly steer agents toward correct bottlenecks, complex trace data might overwhelm or mislead them in certain contexts. While these single-run findings require further replication, they strongly underscore that advancing from repository-level to system-level optimization requires transitioning agents from static code analyzers into runtime-aware profilers.

\subsection{Case Studies}
\label{sec:case_study}

To ground the benchmark mechanics and agent behaviors, we detail three representative instances. Full trajectories, verbatim agent logs, and detailed metrics are available in Appendix~\ref{app:case_study}.

\textbf{Case Study A: Full Recovery.}
On \texttt{martian-bank/nh} (a system injected with both nano-service and
hub-like-dependency anti-patterns), OpenHands + GLM demonstrated precise
architectural diagnosis. The injection splits the transfer path into four
services (transactions $\rightarrow$ sender-resolver $\rightarrow$
receiver-resolver $\rightarrow$ executor) behind a central write entry,
adding four HTTP hops per transfer and degrading latency by 50\%
(9.22\,ms $\rightarrow$ 13.81\,ms). In a 26-minute session (106 actions),
the agent merged the three injected sub-services back into the
transactions service, collapsed per-party account queries into a single
\texttt{\$in} lookup with one \texttt{bulk\_write}, and switched to
persistent HTTP sessions. Latency recovered to 9.04\,ms for a score of
100, landing at the edge of the reference implementation's own
run-to-run range (9.05--9.39\,ms), i.e., convergence to reference-level
performance rather than a demonstrated surpassing of it. This confirms
that co-injected structural patterns are fully repairable within budget, and that the decisive factor is correct localization: the sibling \texttt{n} variant scored 0 under the same configuration in the same batch.

\textbf{Case Study B: Right Fixes, Wrong Bottleneck.} 
On \texttt{media/c2hl} (chatty service combined with hub-like serialization), Claude Code + DeepSeek quickly navigated the 13-service codebase. It successfully merged seven per-field review RPCs into one and stripped legacy lookups, passing both evaluation gates in under nine minutes. However, the agent directly examined the primary defect—a global mutex wrapping five serial cache reads—and explicitly dismissed it in its logs: \emph{``The 5 individual gets are cheap\dots Let me leave this as is.''} Because the agent relied purely on static analysis, it failed to realize that under concurrent traffic, this global lock forces every request into a severe lock convoy, causing the injected $141\times$ latency spike. Latency remained stranded at 5,951.80\,ms (score 0). This case shows the limits of static reasoning. The agent possessed the coding ability to fix the issue, but lacked the dynamic execution data required to recognize its severity.

\textbf{Case Study C: Measurement Reorders Priorities.} 
On \texttt{hotel/c2s}, the baseline Claude Code + Qwen scored 0, 0, and 75.3 across three runs. In the failing runs, the agent used static code reading to mistakenly identify chatty RPCs as the primary bottleneck, leaving a severely serialized shared-DB path unrepaired. However, when equipped with the profiling tool (Section~\ref{sec:exp:profiling}), the agent's behavior transformed. Using actual load measurements and an A/B comparison (22.2\,s vs.\ 10.8\,ms), the agent proved that the shared-DB path was the true culprit. It then fixed both injected patterns in a 14-minute session, achieving a score of 88.8, outperforming its best baseline run. This case cleanly isolates the value of runtime augmentation. It does not grant the agent better coding skills, but it provides the critical evidence needed to align the agent's optimization priorities with reality.

\section{Related Work}

\noindent\textbf{Code Performance Optimization.}
The rapidly growing body of work on LLM code performance
optimization has produced function-level benchmarks~\citep{huang2024effibench,qiu2025efficient,waghjale2024ecco,du2024mercury} and, more recently, repository-level benchmarks~\citep{shetty2026gso,he2025swe,ma2025swe}.
MicroPerf-Bench extends this progression to the system level: rather than optimizing a single codebase against its own workload, agents must recover performance that emerges from interactions among independently running services.

\noindent\textbf{Microservice Benchmarks.}
Prior microservice benchmarks have mainly been used to study system performance, resource efficiency, reliability, testing, and operation~\citep{gan2019open,sriraman2018mu,von2018teastore,zhou2018benchmarking}.
With the development of coding agents, recent work has begun to evaluate end-to-end code generation for microservice systems~\citep{peng2026repogenesis}.
Beyond functional correctness, performance is also important because end-to-end behavior in microservice systems depends on runtime interactions among multiple services.
MicroPerf-Bench focuses on this capability by evaluating whether coding agents can optimize the end-to-end performance of existing microservice systems while preserving their functionality.


\section{Conclusion}
\label{sec:conclusion}

We introduced MicroPerf-Bench, a benchmark for evaluating coding agents on end-to-end performance optimization in microservice systems.
It contains 65 instances across seven systems and nine performance anti-patterns constructed through controlled anti-pattern injection.
Our evaluation of nine agent--model configurations shows that current agents remain far from solving these tasks, with the best configuration achieving an average normalized score of 42.0 out of 100.
Many unsuccessful attempts remain functionally correct but produce no measurable end-to-end performance improvement, highlighting the difficulty of identifying system-level runtime bottlenecks.
Performance also varies across host systems and repeated runs.
Profiling results further suggest that runtime evidence can improve optimization effectiveness.

\section*{AI use statement}

In this work, we did not use generative AI tools for any research tasks requiring
mandatory disclosure, including developing theoretical or conceptual frameworks,
formulating mathematical claims or proofs, proposing or refining hypotheses,
designing research methodology or experiments, implementing research methods,
processing datasets, or interpreting results. Tasks involving translation,
synthetic data generation, and qualitative or thematic data analysis are not
applicable to this work.

We emphasize that all core technical artifacts of this work were produced entirely
by the authors without generative AI assistance. In particular, all anti-pattern
injection code used to construct the benchmark instances, as well as all manual
performance optimizations applied to the original reference implementations, were
designed, written, and validated by human authors.

We used generative AI tools for language editing to improve the readability,
clarity, and fluency of the manuscript, and for limited assistance in writing,
debugging, and troubleshooting simple experimental scripts (e.g., batch execution
and result-aggregation utilities). Such assistance was restricted to suggesting
code snippets, identifying potential coding issues, and improving code readability,
and did not involve the design or implementation of any research method, benchmark
artifact, or evaluation logic.

All AI-assisted content was reviewed and verified by the authors. AI-assisted
language revisions were manually checked to ensure that they preserved the intended
technical meaning and did not introduce unsupported claims, factual changes, or
references. Any AI-assisted code was manually inspected, executed, and tested by
the authors to verify its correctness and consistency with the intended
experimental setup.

We take full responsibility for the final content of this work, including all text,
claims, code, and other artifacts produced with the aid of generative AI.

\section*{Reproducibility Statement}
All benchmark instances, including the under-optimized codebases, Docker Compose configurations, workload scripts, functional tests, and task specifications, together with the evaluation harness used to produce our results, are publicly available at \url{https://github.com/zyzcjnb/microperf}. The exact task specification issued to all evaluated agents is provided in Appendix~\ref{app:taskspec}, and the full evaluation procedure (environment, warm-up, repetitions, and scoring) is described in Section~\ref{sec:exp}. We release the per-run score of all $9 \times 65 \times 3 = 1{,}755$ optimization runs together with the failure-mode label of every zero-score trial, so that all aggregates in this paper can be recomputed and re-aggregated under alternative protocols. The degraded-baseline latency of each instance is constant across all three runs (verified across all 27 result files), which is what makes the repetitions directly comparable.

\section*{Ethics Statement}
The benchmark is constructed from publicly available open-source microservice systems, and their original licenses are respected.
Injected variants are sanitized to mitigate benchmark contamination, as described in Section~\ref{sec:data_construction}.
The experts involved in benchmark construction were recruited through the authors' personal networks and participated voluntarily. We thank them for their time and expertise.



\bibliographystyle{iclr2027_conference}
\bibliography{iclr2027_conference}


\newpage

\appendix

\section{Pre-Study: Mapping Anti-Patterns to Real-World Performance Issues}
\label{app:mapping}

\paragraph{Methodology.}
We mined the commit histories of 12 well-known open-source microservice
systems and catalogued documented industrial performance cases. A commit
qualifies as \emph{system-level} only if it (i) modifies interactions
across at least two services, a shared data/cache tier, or the service
decomposition structure, and (ii) has an explicit performance
motivation. We excluded function-level optimizations within a single
service (e.g., serializer caching), pure parameter tuning (e.g.,
connection limits, GC settings), and dependency upgrades. Two authors
independently screened candidates and resolved disagreements by
discussion.

\paragraph{Mining results.}
Table~\ref{tab:mining} summarizes the outcome. Among more than 10{,}000
commits, only 15 (0.15\%) satisfy our criteria, and they concentrate in
repositories maintained by corporate performance teams. This sparsity
and skew empirically motivate injection-based construction.

\begin{table}[t]
  \centering
  \caption{Pre-study: mining real-world system-level performance fixes
  from open-source microservice systems.}
  \label{tab:mining}
  \small
  \begin{tabular}{lrr}
    \toprule
    Repository & Commits & Qualifying fixes \\
    \midrule
    DeathStarBench (3 applications)      & 599   & 9 \\
    eShopOnContainers (Microsoft)        & 4{,}339 & 1 \\
    OpenTelemetry Demo (CNCF)            & 2{,}077 & 0 \\
    TeaStore (academic benchmark)        & 1{,}766 & 1 \\
    Spring PetClinic Microservices       & 1{,}500$^{\ast}$ & 2 \\
    Robot Shop (Instana)                 & 377   & 0 \\
    TrainTicket (academic benchmark)     & 323   & 0 \\
    Online Boutique (Google)             & ---$^{\dagger}$ & 1 \\
    AcmeAir (IBM)                        & 77    & 0 \\
    MartianBank (Cisco)                  & 48    & 0 \\
    Pitstop                              & ---$^{\dagger}$ & 1 \\
    FTGO / PiggyMetrics / Sock Shop      & issue-level scan & 0 \\
    \midrule
    \textbf{Total}                       & $>$10{,}000 & \textbf{15} \\
    \bottomrule
  \multicolumn{3}{l}{\scriptsize $^{\ast}$~Approximate; mined via API.
  $^{\dagger}$~Partial history; mined via API and issue search.}
  \end{tabular}
\end{table}


\paragraph{Mapping to the nine anti-patterns.}
Table~\ref{tab:mapping} maps each injected anti-pattern to concrete
real-world evidence. Seven of the nine anti-patterns are directly
corroborated by merged commits in open-source systems or by documented
industrial cases. The remaining two, \emph{Empty messages} and
\emph{ESB misuse}, are anchored in the empirical anti-pattern
catalog~\citep{cerny2023catalog}: Empty messages traces back to the
original web-service anti-pattern literature~\citep{rodriguez2010improving},
where it has been empirically detected and refactored in real
services~\citep{ordialescoscia2013antipattern}, while ESB misuse is
corroborated by its modern counterpart, the removal of Istio's
centralized Mixer component. We also observed that two major demo
systems (OpenTelemetry Demo, Robot Shop) ship intentionally injected
performance faults, indicating that fault injection is the community's
prevailing practice when a performance defective microservice system is
needed.

\begin{table}[t]
  \centering
  \caption{Mapping injected anti-patterns to real-world evidence.
  \textbf{C}~=~merged commit(s) in open-source systems;
  \textbf{I}~=~documented industrial case; \textbf{L}~=~empirical
  literature catalog.}
  \label{tab:mapping}
  \footnotesize
  \begin{tabularx}{\linewidth}{@{}l >{\raggedright\arraybackslash}p{3.2cm} >{\raggedright\arraybackslash}X c@{}}
    \toprule
    Anti-pattern & System / Source & Real-world evidence & Type \\
    \midrule
    Chatty service &
      PetClinic PR\,\#95; DSB \texttt{ed98db1}; eShop \texttt{2d59b7cc} &
      Batch endpoints replace per-item cross-service calls; memcached
      multi-get replaces per-key loops; per-request gRPC clients replaced
      by pooled factories & C \\
    & SoundCloud (BFF); Meta (GraphQL) &
      Fine-grained multi-round calls and over-fetching motivated the
      BFF pattern and GraphQL & I \\
    Shared persistency &
      DSB PRs \#268/\#271/\#206; Redis replication series &
      Multiple memcached backends, mcrouter, and Redis replication added
      to relieve shared cache-tier contention & C \\
    & Discord messaging storage &
      Hot partitions degraded unrelated queries; request coalescing and
      routing reduced p99 from 40--125\,ms to $\sim$15\,ms & I \\
    Service chain &
      TeaStore pool series; TrainTicket industrial faults &
      Cross-service REST pool and load-balancer timeout fixes; recursive
      cross-service request amplification replicated from industrial
      faults~\citep{deeptralog2022} & C/L \\
    Hub-like dependency &
      Istio Mixer removal &
      Centralized telemetry/policy hub on the request path removed;
      p90 latency halved, CPU per 1k requests reduced by 50\% & I \\
    Nano-service / Megaservice &
      Segment; Istio istiod; Amazon Prime Video &
      Public consolidations of over-decomposed services, motivated by
      head-of-line blocking and cross-component communication overhead & I \\
    On-line only &
      Netflix Zuul 2; PetClinic WebFlux migration &
      Blocking gateway rewritten to async non-blocking I/O; blocking
      REST clients replaced by reactive clients & I/C \\
        Empty messages &
      Anti-pattern catalogs; real web services &
      A message that omits business data forces the receiver to query its
      own storage or another service to retrieve the missing information;
      cataloged anti-pattern~\citep{cerny2023catalog} originating from the
      web-service literature~\citep{rodriguez2010improving}, where it was
      detected and refactored in 154 real services~\citep{ordialescoscia2013antipattern} & L \\
    ESB misuse &
      Istio Mixer (modern counterpart) &
      Central hub on the data path identified as the system bottleneck & I/L \\
    \bottomrule
  \end{tabularx}
\end{table}

\paragraph{Limitations.}
Naturally occurring fixes rarely ship with measurements or reproducible
environments (e.g., commit \texttt{ed98db1} carries a two-word message
with no linked issue), which is precisely why they cannot serve directly
as benchmark instances. Some anti-patterns (e.g., \emph{Empty messages})
manifest as persistent design choices rather than one-shot fixes, so
their evidence is industrial rather than commit-level. Finally, injected
instances may not capture the full complexity of open-ended, organically
emerging performance problems; we mitigate this threat by deriving all
operators from an empirically grounded catalog and by corroborating each
operator with the evidence above.

\section{Task Specification Provided to Agents}
\label{app:taskspec}

For reproducibility, we provide the exact system prompt and user
instruction issued to every evaluated agent. The prompt defines the
optimization objective, the three permitted modification levels (code
logic, service architecture, and service topology), and the fairness
constraints enforced by the gates in Section~\ref{sec:bench_eval}. All agents receive
an identical specification; only the placeholder \texttt{\{work\_dir\}}
is instantiated with the instance-specific project path.


\newpage

\noindent\textbf{System prompt.}
\begin{lstlisting}[style=promptbox]
You are a microservice performance optimization expert.

The current directory contains a microservice system with potential performance issues.

These issues may stem from code logic, service topology, or the overall architecture.

Your task is to independently analyze and optimize this system to improve its performance (reduce latency, increase throughput, and enhance scalability).

Requirements:

1. Keep the original API interfaces and basic functionality unchanged.

2. Use tools to read, analyze, and modify the code files.

3. After modification, ensure that the optimized code files are written back to the current directory.

4. Optimization must be achieved ONLY at three levels: (a) code logic, (b) service architecture, (c) service topology. When architectural changes are required, such as merging or splitting services, or changing communication patterns, you can modify business logic code (e.g., server.go, main.go, db.go, server.js, *.py, etc.) and service topology/deployment files (e.g., docker-compose.yml, Dockerfiles) - but ONLY to change the service architecture/topology itself.

5. Do not improve performance through resource-level changes: Do not increase hardware limits such as CPU/memory, increase the number of replicas, or add extra instances of existing services.

6. Do not modify generated files (e.g., proto/*.pb.go, vendor/, node_modules/) or scripts used for testing.

7. Focus on the real bottlenecks; avoid unnecessary rewrites.

8. Performance gains must come from eliminating real inefficiencies in code, topology, or architecture - not from making the system do less work than it is supposed to do. Do NOT weaken or disable non-functional safeguards to gain performance, including: distributed tracing/observability (e.g., do not disable Jaeger or lower its sampling rate), logging, data durability mechanisms (e.g., MongoDB journaling, Redis AOF/snapshot persistence, fsync), security mechanisms, and correctness-related semantics (e.g., ordering guarantees, error propagation).

9. Do NOT perform any parameter or configuration tuning. Do not modify configuration files or settings, including but not limited to: nginx.conf and other *.conf files, *-config.json / *.yml settings, database startup flags and command options, keepalive intervals, connection pool sizes, worker/thread counts, sampling rates, buffer/cache sizes, and OS or runtime parameters. docker-compose.yml and Dockerfiles may be modified ONLY for architectural changes (merging, splitting, adding, or removing services, or changing communication patterns); image names, container command lines, environment variables, and resource settings must otherwise remain untouched.

Current working directory: {work_dir}
\end{lstlisting}

\noindent\textbf{User instruction.}
\begin{lstlisting}[style=promptbox]
Please analyze the microservice system in the current directory ({work_dir}) and optimize its performance by modifying the code.

The steps are as follows:

1. List and read the key files in the microservice system: business logic, service topology, and communication patterns.

2. Identify performance bottlenecks at the code, service topology, and architecture levels.

3. You may modify docker-compose.yml and Dockerfiles ONLY to change the service architecture or topology (e.g., merging or splitting services); do not increase hardware limits such as CPU/memory, do not increase the number of replicas or add extra instances of existing services, and do not tune any parameters or configuration settings.

4. Write the optimized code back to the original files.

5. Report which files were modified and what was changed.

Start optimizing now!
\end{lstlisting}

\smallskip
\noindent The two modules above constitute the complete task specification
issued to all evaluated agents. The placeholder \texttt{\{work\_dir\}} is
replaced with the instance-specific project path at runtime.

\section{Expert Validation}
\label{app:expert_validation}

\paragraph{Reviewers and Procedure.}
We recruited three experts with over three years of experience in microservice development through the authors' personal networks.
All experts participated voluntarily.
Following anti-pattern injection, the resulting benchmark instances were subjected to expert validation before being included in the final benchmark.
The validation consisted of two complementary assessments: anti-pattern diagnosis and defect-source discrimination.
The former examines whether the injected performance problems exhibit identifiable runtime symptoms, while the latter examines whether the injected instances contain obvious cues that reveal their artificial origin.

\paragraph{Anti-Pattern Diagnosis.}
All 65 benchmark instances were included in the diagnosis assessment.
For each instance, reviewers were asked to identify the injected performance anti-pattern from its runtime behavior without being provided with the corresponding anti-pattern label.
The injected anti-pattern was correctly diagnosed in $81.5\%$ of cases.
This result indicates that the injected performance problems generally produced recognizable runtime symptoms under the evaluation setting.
We use this assessment to evaluate the diagnosability of the constructed instances rather than to claim equivalence between injected and naturally occurring performance defects.

\paragraph{Defect-Source Discrimination.}
We further conducted a blinded discrimination study to examine whether the injected instances exhibited obvious artifacts of the injection process.
We sampled 12 injected instances and mixed them with 12 naturally occurring performance defects reconstructed from the parent commits of documented performance fixes in open-source repositories.
The 24 cases were presented without revealing whether their source was injected or naturally occurring.
The reviewers were asked to identify the source of each defect.
The resulting source-identification accuracy was $47.2\%$.
We report this result descriptively: under this evaluation setting, the reviewers did not reliably identify the provenance of the sampled defects, suggesting that the injected instances did not present obvious source-specific cues to the reviewers.

\paragraph{Scope of the Validation.}
These assessments serve as complementary checks on the benchmark construction process.
The diagnosis assessment evaluates whether the injected problems manifest as identifiable performance symptoms, whereas the discrimination assessment evaluates whether their injected provenance is readily apparent to expert reviewers.
The latter does not constitute a statistical equivalence test between injected and naturally occurring defects, nor does it establish that the two groups have identical runtime behavior or optimization difficulty.
Instead, it provides evidence that the sampled injected instances do not exhibit obvious artificial characteristics under the expert evaluation setting.

\section{Agent Configurations}
\label{app:agent_config}

Table~\ref{tab:agent_config} summarizes the versions, runtimes, and invocation
modes of the three agent scaffolds. To ensure a controlled comparison, all
scaffolds access the same three base models (GLM-5.2, Qwen3.7-Max, and
DeepSeek-V4-Pro) through the same Alibaba Cloud token-plan gateway, so that
identical model endpoints are served to every scaffold without vendor-specific
coupling. All models are run with temperature $T = 1.0$ and thinking mode
enabled. Each task instance is optimized in an independent, scripted,
non-interactive invocation; no interactive sessions or manual interventions are
involved at any point.

\begin{table}[h]
  \centering
  \small
  \caption{Agent scaffold versions and configurations. All three scaffolds are
  driven in scripted, non-interactive (headless) mode, with one independent
  invocation per benchmark instance.}
  \label{tab:agent_config}
  \setlength{\tabcolsep}{4pt}
  \begin{tabular}{llp{5.2cm}p{4.6cm}}
    \toprule
    Scaffold & Version & Model access & Invocation \\
    \midrule
    OpenCode & 1.18.3 &
    Alibaba Cloud token-plan gateway &
    \texttt{opencode run}, one scripted call per instance \\
    OpenHands & 1.11.0 &
    Alibaba Cloud token-plan gateway &
    Headless CLI mode, one scripted call per instance \\
    Claude Code & 2.1.114 &
    Alibaba Cloud token-plan gateway
     &
    Non-interactive headless mode, one scripted call per instance \\
    \bottomrule
  \end{tabular}
\end{table}

All scaffolds receive an identical task specification (Appendix~\ref{app:taskspec}),
an identical containerized execution environment, and a unified 2-hour time budget
per instance with a 20-minute timeout for each individual command or tool call.
Agents are allowed to iterate freely within a single optimization run, but we never
relaunch an agent on the output of a previous run, so the three repetitions of each
configuration are independent draws from the same configuration-level distribution.

\section{Case Studies}
\label{app:case_study}

To make the benchmark mechanics, agent behavior, and scoring concrete, we examine
three representative instances in detail (Table~\ref{tab:case_study}): a full-score
repair of a combined injection (Case~A), a gate-compliant run that achieved no
improvement (Case~B), and a profiling-guided turnaround by Claude Code + Qwen
(Case~C).

\begin{table}[h]
  \centering
  \small
  \caption{Summary of the three case-study instances. $L_R$, $L_D$, and $L_O$ are
  the average end-to-end latencies of the reference, degraded, and optimized
  versions under the standard mixed workload.}
  \label{tab:case_study}
  \setlength{\tabcolsep}{4pt}
  \resizebox{\textwidth}{!}{%
  \begin{tabular}{llllrrrr}
    \toprule
    Case & Configuration & Instance & Injected patterns & $L_R$ & $L_D$ & $L_O$ & Score \\
    \midrule
    A & OpenHands + GLM & martian-bank / nh & Nano-service + Hub-like dep. &
9.22 & 13.81 & 9.04 & 100.0 \\
    B & Claude Code + DeepSeek & media / c2hl & Chatty service + Hub-like dep. &
30.13 & 4260.99 & 5951.80 & 0.0 \\
    C & Claude Code + Qwen & hotel / c2s & Chatty service + Shared persistency & 3.78 & 42.07 & 8.05 & 88.8 \\
    \bottomrule
      \end{tabular}}
\end{table}

\textbf{Case A: full recovery of a combined structural injection.} The
nano-service injection re-splits the original local transfer logic into
a chain of three fine-grained services (sender-resolver,
receiver-resolver, executor) behind the transactions service, and the
hub-like-dependency injection routes all balance updates through a
central \texttt{/update-balance} entry in \texttt{accounts}, so a single
transfer incurs four additional HTTP hops; the whole injection touches
only four files and adds three service directories with zero functional
change. The reference runs at 9.22\,ms and the injected variant at
13.81\,ms (+50\%). In 1,564 seconds (106 actions: 53 terminal commands,
44 file edits), the agent removed the three injected service definitions
from \texttt{docker-compose.yaml}, merged the resolver/executor logic
back into \texttt{transactions/transaction.py} with direct MongoDB
access (one \texttt{\$in} lookup for both parties, a single
\texttt{bulk\_write} for balances), and replaced per-request HTTP
connections with module-level \texttt{requests.Session} in
\texttt{dashboard/dashboard.py} and \texttt{loan/loan.py}. Its own
bottleneck analysis matches the injected structure almost exactly:
\emph{``The transaction flow is overly chatty\dots Merging the 3
sub-services into the transactions service is the high-impact,
clearly-architectural optimization.''} The evaluation measured an
average latency of 9.04\,ms, an error rate of 0.13\%, and a throughput
of 25.4\,req/s, giving
$(13.81-9.04)/(13.81-9.22)=103.9$, clamped to 100. The optimized
latency sits at the lower edge of the reference implementation's own
run-to-run range (9.05--9.39\,ms), so we report convergence to
reference-level performance within measurement noise, not a
demonstrated improvement beyond the reference; the query batching and
persistent sessions are retained optimizations whose individual effects
are below what a single evaluation can resolve. The agent deliberately
left the injected forwarding path in the loan service untouched, since
the workload exercises transfers, so the recovery is complete for the
measured path. Under the same agent, model, and time window, the
sibling \texttt{n} variant (nano only) scored 0 at 47.68\,ms with 944
failed requests. This case shows that a combined nano + hub injection
is fully actionable by topology restoration, that reference-level
performance is reachable within a small fraction of the 2-hour budget,
and that localization quality rather than model capacity separates the
full recovery from the zero score.

\textbf{Case B: right fixes, wrong bottleneck.} This instance co-injects two patterns into the media service: a chatty-service
injection that turns one batched \emph{memcached\_mget} into five serial
\emph{memcached\_get} calls and one full StoreReview RPC into seven per-field
RPCs, and a hub-like dependency realized as a global
\texttt{std::recursive\_mutex} (added via the Thrift IDL) whose
\texttt{lock\_guard} wraps the memcached fetch section, so every request holds
a global lock across five serial round trips. The result is a severe combined
degradation: reference 30.13 ms, injected 4,260.99 ms ($\sim$141$\times$).
In a nine-minute session (128 assistant messages, 72 tool calls, 801,809
tokens), the agent reads nearly all of the 13-service codebase, correctly
identifies the secondary bottleneck (``It should send the entire review in one
call''), and shows rule awareness (declining to modify the wrk2 test scripts).
It then explicitly identifies the dominant defect and dismisses it: ``The
\emph{\_ComposeAndUpload} function makes 5 separate \emph{memcached\_get} calls.
These could be batched with mget. But these are within a mutex lock\dots The 5
individual gets are cheap (local memcached) compared to the network calls to
other services. Let me leave this as is.'' Its edits are substantial and largely valid (six files, twelve edits): the
seven StoreReview RPCs are merged back into one, legacy lookups are removed,
serial fetches are batched and parallelized, and two zcard checks on pre-existing
original code are removed. Both gates pass and throughput is healthy
($\sim$4,215 requests/s), yet measured latency is 5,951.80 ms against the
4,260.99 ms degraded baseline (a 39.7\% regression), and the score is 0. The failure is not capability but prioritization without measurement. The
dominant cost is the lock convoy, not the reads themselves: under the global
mutex, each in-lock round trip is multiplied by queueing from all concurrent
requests. The bottleneck is demonstrably fixable within the rules: nineteen
other runs of this variant, spanning all nine configurations, recover it to
scores between 58.6 and 100, and the agent's edit set lacks exactly the one step
these recoveries share, restoring the batched read; another run of the same
configuration reverses the ``leave this as is'' decision and scores 100. Nothing in the trajectory flags the run as a failure; only the
measurement does.

\textbf{Case C: measurement changes what the agent optimizes.} Claude Code + Qwen
faced a chatty-service injection (the frontend calls the profile
service once per hotel ID, sequentially) combined with a shared-persistency
injection that merges six per-service MongoDB instances into one shared
instance guarded by per-service mutexes, per-request audit writes, and
discarded-result full-collection counts. Its three baseline runs scored 0, 0, and 75.3, a vivid instance of run-to-run instability, and the two failures are instructive:
reasoning purely from static code, the first run ranked the chatty RPCs as the
top bottleneck (\emph{``Optimization 1: Frontend --- Batch profile requests''}) and
treated the per-request full-collection scan as a normal cache-refill path, leaving
the shared-DB serialization unrepaired; the second rewrote the scan into a locked
single pass and left all other service mutexes in place, collapsing throughput to
4.3\,req/s. With the profiling tool (which enforces a strict budget of two
measurement cycles and surfaces workload measurements and, where deployed, tracing
data), the same configuration re-prioritized its targets: load measurements showing
collapse at 2{,}000\,RPS and an A/B comparison of 22.2\,s vs.\ 10.8\,ms established
\emph{``Global dbMutex serialization\ldots{} Full collection scan in rate
service''} as the primary bottlenecks, demoting the chatty RPCs to one of seven
hypotheses. The final session (14 minutes) fixed both the shared-DB path and the
N+1 calls, reaching 8.05\,ms and a score of 88.8, above the best baseline run.
The contrast isolates what the profiling tool buys: not more editing ability, but
evidence that reorders optimization priorities toward the bottlenecks that actually
dominate end-to-end latency.

\section{Results by Anti-Pattern}
\label{app:antipattern}

Table~\ref{tab:app_antipattern} breaks down optimization performance by the injected
anti-pattern. Combined instances contribute to each of their constituent patterns.
The nine patterns differ markedly in difficulty. Hub-like dependency is the hardest
pattern on average (21.6 points, 8.3\% full-score rate): the injected hub serializes
request traffic, and agents frequently relieve only part of the constraint or migrate
the serialization point rather than removing it. Its difficulty is also
system-dependent, ranging from 4.3 points on Martian and Social to 56.6 on Pitstop,
depending on whether the hub sits on the critical path of the mixed workload.
On-line only and Chatty service are the most tractable patterns (47.9 and 42.5
points), as both have well-known architectural fixes (asynchronous I/O and batching).
Notably, failures on On-line only are dominated by correctness errors (43.3\% of its
zero-score trials): rewriting blocking call chains into asynchronous ones often breaks
ordering or error-propagation semantics. The single Mega service instance (on Social)
was never solved: merging more than ten services broke the build in 51.9\% of its
zero-score trials, indicating that large-scale structural consolidation is infeasible
within the two-hour budget. Empty messages shows the highest positive-score rate
(61.7\%) but a modest average score (34.9), since payload shaping yields only limited
latency gains under our non-saturated workloads; 87.1\% of its zero-score trials are
no-gain outcomes.

\begin{table}[t]
\centering
\small
\caption{Results by injected anti-pattern. Instances denotes the number of benchmark
instances containing the pattern (combined instances count toward each constituent
pattern). Rates use all trials on those instances as the denominator. The last column
gives the dominant failure mode among zero-score trials.}
\label{tab:app_antipattern}
\begin{tabular}{lccccc}
\toprule
Anti-pattern & Instances & Avg.\ score & Positive (\%) & Full score (\%) & Dominant failure (\%) \\
\midrule
Chatty service        & 24 & 42.5 & 50.8 & 23.0 & No perf.\ gain (50.8) \\
Nano-service          & 21 & 38.7 & 46.7 & 19.6 & No perf.\ gain (50.3) \\
Hub-like dependency   & 17 & 21.6 & 31.2 &  8.3 & No perf.\ gain (54.7) \\
Service chain         & 11 & 31.3 & 45.5 &  9.8 & Correctness error (38.9) \\
On-line only          & 10 & 47.9 & 55.6 & 24.8 & Correctness error (43.3) \\
Shared persistency    & 10 & 34.9 & 44.1 & 18.1 & No perf.\ gain (51.7) \\
Empty messages        &  3 & 34.9 & 61.7 & 21.0 & No perf.\ gain (87.1) \\
Mega service          &  1 &  0.0 &  0.0 &  0.0 & Build failure (51.9) \\
ESB misuse            &  0$^{*}$ & -- & -- & -- & -- \\
\bottomrule
\end{tabular}
\\[2pt]
\begin{minipage}{\linewidth}
\footnotesize $^{*}$ESB misuse has no injected instances: no suitable injection target
exists in the selected systems, as none of the widely used open-source microservice
systems contains an ESB component.
\end{minipage}
\end{table}

\section{Failure Analysis Details}
\label{app:failure_detail}

This section extends Section~\ref{sec:exp:failure} with the full six-way breakdown. In the main text,
build failures and deploy/startup failures are reported jointly as build-or-startup
failures (14.4\% of zero-score trials); here we report them separately.
Table~\ref{tab:app_failure_overall} gives the overall distribution. Beyond the dominant
no-gain category (48.7\%), correctness errors (20.5\%) and fairness violations (15.3\%)
account for most remaining failures, while timeouts are rare (1.1\%).

\begin{table}[t]
\centering
\small
\caption{Distribution of the 982 zero-score trials across six failure modes.}
\label{tab:app_failure_overall}
\begin{tabular}{lcc}
\toprule
Failure mode & Count & Share (\%) \\
\midrule
No performance gain      & 478 & 48.7 \\
Correctness error        & 201 & 20.5 \\
Fairness violation       & 150 & 15.3 \\
Build failure            & 111 & 11.3 \\
Deploy / startup failure &  31 &  3.2 \\
Timeout                  &  11 &  1.1 \\
\bottomrule
\end{tabular}
\end{table}

\begin{table}[t]
\centering
\footnotesize
\caption{Failure-mode composition by system. Mode shares are computed within each
system's zero-score trials. NG: no performance gain; FV: fairness violation; CE:
correctness error; BF: build failure; DS: deploy/startup failure; TO: timeout.}
\label{tab:app_failure_system}
\begin{tabular}{lcccccccccc}
\toprule
System & Trials & Zero & Zero (\%) & Avg.\ score & NG & FV & CE & BF & DS & TO \\
\midrule
Social    & 297 & 260 & 87.5 &  7.4 & 34.2 &  0.4 & 44.2 & 21.2 &  0.0 & 0.0 \\
RobotShop & 270 & 195 & 72.2 & 11.1 & 67.7 & 22.6 &  3.1 &  5.6 &  1.0 & 0.0 \\
Martian   & 216 & 194 & 89.8 &  6.1 & 46.9 & 48.5 &  2.1 &  0.5 &  1.0 & 1.0 \\
PetClinic & 243 &  87 & 35.8 & 57.5 & 11.5 &  0.0 & 48.3 & 13.8 & 26.4 & 0.0 \\
Pitstop   & 270 &  85 & 31.5 & 54.7 & 80.0 &  1.2 &  2.4 & 10.6 &  0.0 & 5.9 \\
Media     & 189 &  72 & 38.1 & 50.6 & 43.1 &  8.3 & 15.3 & 25.0 &  2.8 & 5.6 \\
Hotel     & 270 &  89 & 33.0 & 60.4 & 64.0 &  4.5 & 23.6 &  5.6 &  2.2 & 0.0 \\
\bottomrule
\end{tabular}
\end{table}

Table~\ref{tab:app_failure_system} shows that failure structure is largely determined
by the system rather than by the configuration. Martian's failures are almost evenly
split between no-gain outcomes (46.9\%) and fairness violations (48.5\%); the latter
mostly come from agents removing simulated-latency sleeps,
which the audit rejects. PetClinic is the only system with a substantial
deploy/startup share (26.4\%), consistent with its slow JVM startup frequently
exceeding the gateway readiness window after modification. Social combines correctness
errors (44.2\%) with build failures (21.2\%), reflecting its large polyglot codebase.
RobotShop and Pitstop failures are dominated by no-gain outcomes (67.7\% and 80.0\%):
agents produce valid, compliant changes that do not move end-to-end latency.

\begin{figure}[t]
\centering
\includegraphics[width=\linewidth]{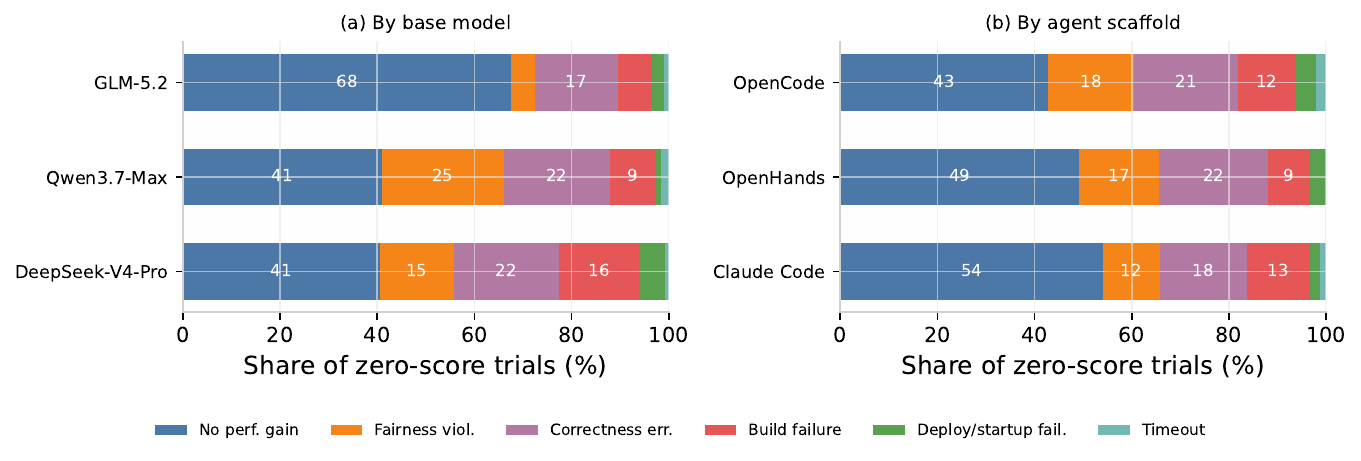}
\caption{Failure-mode composition of zero-score trials (a) by base model and (b) by
agent scaffold. Percentages are computed within each group's zero-score trials; in-bar
labels are shown for segments of at least 7\%.}
\label{fig:app_failure_model}
\end{figure}

Figure~\ref{fig:app_failure_model} aggregates failures by base model and by scaffold.
Model identity shapes the failure profile more than the scaffold. GLM-5.2 is the most
conservative model: 67.6\% of its failures are no-gain outcomes and only 4.8\% are
fairness violations. Qwen3.7-Max is the most aggressive, with the highest fairness
share (25.1\%), mostly prohibited parameter tuning and removal of simulated work.
DeepSeek-V4-Pro produces the most build and deploy/startup failures (16.4\% and 5.3\%),
consistent with its lowest average score. Scaffold differences are smaller; the most
visible one is that OpenHands produced no timeout in any zero-score trial, while
OpenCode accounts for most timeouts (2.1\%).

\section{Per-System Scores by Configuration}
\label{app:heatmap}

\begin{figure}[t]
\centering
\includegraphics[width=\linewidth]{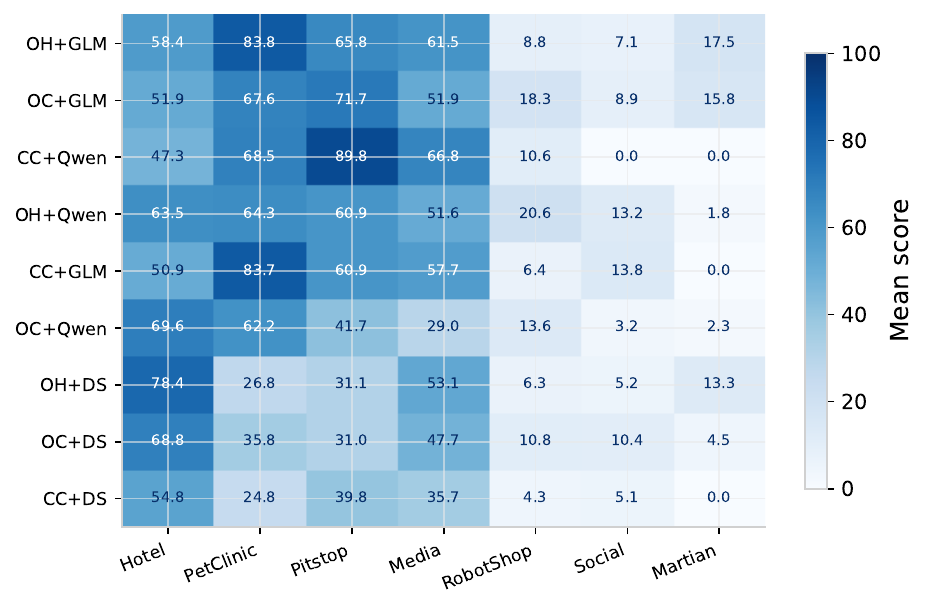}
\caption{Mean score of each configuration on each system (three runs, 65 instances).
Rows and columns are sorted by average score.}
\label{fig:app_heatmap}
\end{figure}

Figure~\ref{fig:app_heatmap} shows the mean score of every configuration on every
system. Two observations stand out. First, system difficulty is remarkably consistent
across configurations: Martian, Social, and RobotShop are hard for every configuration
(no cell above 21 points), whereas Hotel, PetClinic, and Pitstop are tractable for most
configurations. Second, no configuration dominates everywhere. Claude Code + Qwen
reaches the highest single cell (89.8 on Pitstop) but scores 0.0 on Social; OpenHands +
DeepSeek leads on Hotel (78.4) but drops to 6.3 on RobotShop; OpenHands + GLM, the best
overall configuration, is the most balanced one and the only configuration without a
near-zero system (minimum 7.1 on Social). This complementarity suggests that scaffold
and model choices interact with system characteristics such as language stack, service
count, and where the injected degradation sits on the request path.

\section{Single-Pattern vs.\ Combined-Pattern Instances}
\label{app:single_combined}

\begin{figure}[t]
\centering
\includegraphics[width=\linewidth]{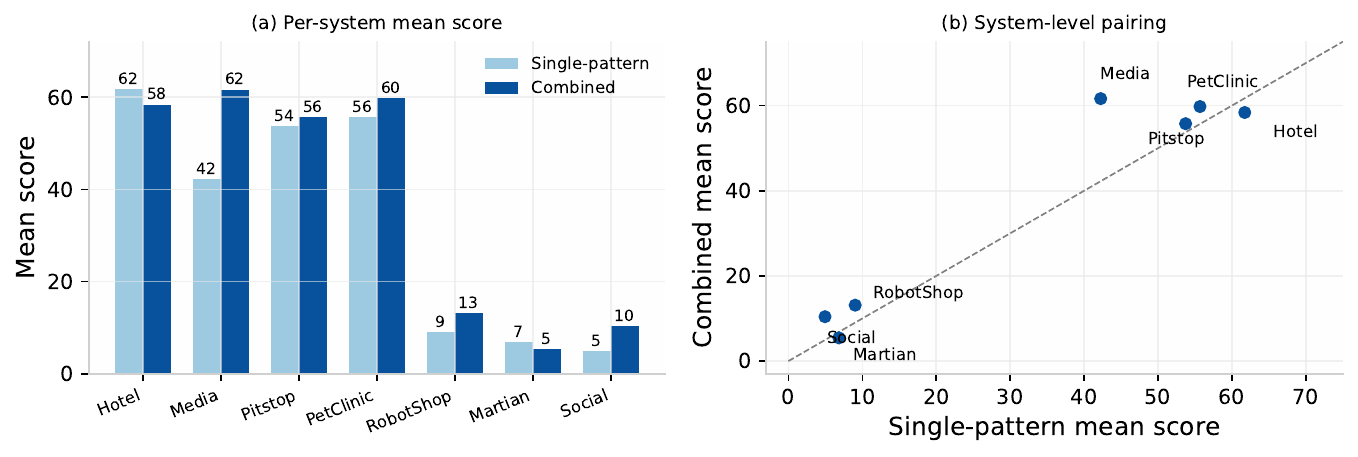}
\caption{Single-pattern versus combined-pattern instances. (a) Per-system mean scores
on the two subsets. (b) System-level pairing; points above the diagonal indicate
systems where combined instances score higher.}
\label{fig:app_single_combined}
\end{figure}

The benchmark contains 35 single-pattern instances and 30 combined instances (28 with
two patterns, 2 with three patterns). Contrary to the intuition that combined instances
should be harder, agents score slightly higher on them: 35.9 versus 34.0 on average,
with positive-score rates of 46.7\% versus 41.8\% and similar full-score rates
(18.0\% versus 15.6\%). A paired comparison within the seven systems confirms the
pattern: combined instances score higher in five of seven systems, with a mean
within-system difference of +4.3 points that does not reach significance (paired
$t(6)=1.56$, $p=0.170$). Per-system scores on the two subsets are strongly correlated
(Pearson $r=0.96$), that is, system difficulty transfers almost perfectly between
single and combined injections (Figure~\ref{fig:app_single_combined}). One plausible
explanation is that co-injected patterns offer multiple independent opportunities for
partial recovery and more diagnostic symptoms, so fixing either pattern already yields
a positive score. Media shows the largest gap (+19.4 points), where its combined
instances pair the two most tractable patterns. We note that the comparison is
confounded by which patterns were selected for combination, and report it as a
descriptive property of the benchmark rather than a causal claim.

\section{Run-to-Run Stability}
\label{app:stability}

Table~\ref{tab:app_stability} quantifies the stochasticity of agentic optimization
across the three independent runs of each configuration. Run-level standard deviations
range from 1.0 (Claude Code + GLM) to 5.9 (OpenCode + GLM), and between 23.1\% and
50.8\% of instances are flaky, meaning that the same configuration produces both zero
and positive scores across runs on the same instance. The gap between the mean score
and Best@3 (the per-instance maximum across the three runs, averaged) ranges from 12.3
to 25.7 points; pooled across configurations, Best@3 reaches 53.7 compared with the
overall mean of 34.8. Claude Code + Qwen is the most stable high-scoring configuration
(SD 1.3, 23.1\% flaky), while OpenCode + GLM combines high variance (SD 5.9, 49.2\%
flaky) with the highest capability ceiling (Best@3 of 65.9). These results show that
single-run evaluations would substantially underestimate capability and can even
reverse configuration rankings, which motivates our three-run protocol and the use of
per-instance paired tests in Section~\ref{sec:exp:factors}.


\begin{table}[t]
\centering
\footnotesize
\setlength{\tabcolsep}{4.5pt}
\caption{Run-to-run stability of the nine configurations (three independent runs,
65 instances each). Range is the spread between the highest and lowest per-run
mean. Best@3 averages the per-instance maximum score across the three runs;
Uplift is the gap between Best@3 and the mean. Flaky denotes the share of
instances with both zero and positive scores across runs.}
\label{tab:app_stability}
\begin{tabular}{llrrrrr}
\toprule
Scaffold & Model & Mean & SD & Range & Best@3 (uplift) & Flaky (\%) \\
\midrule
OpenHands   & GLM      & 42.0 & 5.3 & 10.0 & 60.6 (+18.6) & 41.5 \\
OpenCode    & GLM      & 40.2 & 5.9 & 11.8 & 65.9 (+25.7) & 49.2 \\
Claude Code & Qwen     & 39.4 & 1.3 &  2.4 & 51.7 (+12.3) & 23.1 \\
OpenHands   & Qwen     & 39.2 & 1.5 &  2.9 & 61.6 (+22.4) & 35.4 \\
Claude Code & GLM      & 38.3 & 1.0 &  1.9 & 51.1 (+12.8) & 32.3 \\
OpenCode    & Qwen     & 31.8 & 5.0 &  9.9 & 52.2 (+20.4) & 50.8 \\
OpenHands   & DeepSeek & 29.8 & 1.6 &  3.2 & 47.2 (+17.4) & 38.5 \\
OpenCode    & DeepSeek & 29.4 & 2.5 &  5.0 & 51.8 (+22.4) & 44.6 \\
Claude Code & DeepSeek & 23.4 & 3.7 &  6.7 & 41.5 (+18.1) & 40.0 \\
\bottomrule
\end{tabular}
\end{table}

\section{Cost and Time Analysis}\label{app:cost}

Beyond effectiveness, practical adoption of coding agents depends on cost
and latency. We therefore quantify the resource footprint of the main
evaluation. All configurations are served through the same Alibaba Cloud
token-plan gateway (Appendix~\ref{app:agent_config}), so per-token prices are identical across
configurations and scaffolds. We report billing-equivalent tokens per
instance, computed as $4\times$ output $+$ $1\times$ uncached input $+$
$0.1\times$ cache-read tokens under the plan's standard weights. Raw token
counts are dominated by cache reads (94.3\% on average, range 88.2--97.6\%),
which reflect scaffold context management rather than newly generated
content. Aggregated over all $9\times65\times3=1{,}755$ trials, the main
evaluation consumed 8.13B raw tokens (1.38B billing-equivalent) and roughly
369 agent-hours of wall-clock time.

\begin{figure}[t]
\centering
\includegraphics[width=\textwidth]{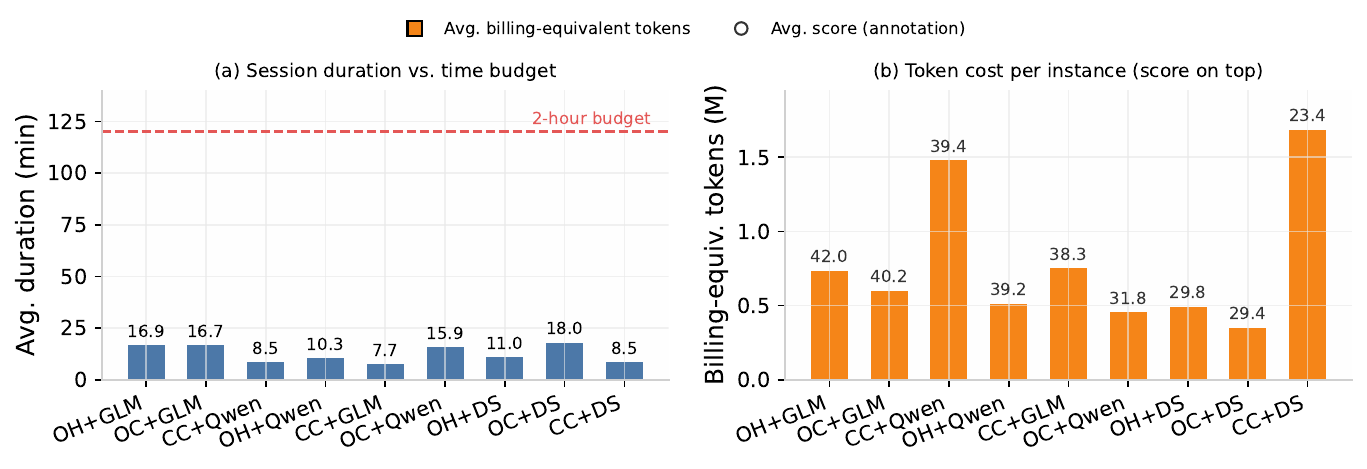}
\caption{Cost and time efficiency of the nine configurations, ordered by
average score. (a) Mean session duration against the two-hour budget
(dashed line). All configurations use less than 15\% of the budget.
(b) Billing-equivalent tokens per instance; the annotation above each bar
is the configuration's average score. Token cost is uncorrelated with
effectiveness: Claude Code + DeepSeek is the most expensive configuration
and the worst-scoring one.}
\label{fig:cost}
\end{figure}

Table~\ref{tab:cost} and Figure~\ref{fig:cost} break cost and duration down
by configuration. Token cost varies widely, from 0.35M billing-equivalent
tokens per instance (OpenCode + DeepSeek) to 1.69M (Claude Code + DeepSeek).
The scaffold, rather than the model, is the primary driver: averaged over
models, Claude Code consumes $2.8\times$ more tokens than OpenCode
(1.31M vs.\ 0.47M) and $2.3\times$ more than OpenHands (0.58M), while model
differences are smaller (GLM 0.70M, Qwen 0.81M, DeepSeek 0.84M on average).
Cost composition also differs: cache reads account for 72.5\% of OpenCode +
GLM's billing-equivalent tokens but only 43.6\% for OpenHands + Qwen,
reflecting different context-reuse strategies.

Crucially, cost does not buy performance. The correlation between
per-instance token cost and average score across the nine configurations is
negligible (Pearson $r=-0.20$, $p=0.60$), and score-per-cost varies by
$6\times$ (84.2 score points per 1M billing-equivalent tokens for OpenCode +
DeepSeek vs.\ 13.9 for Claude Code + DeepSeek, the most expensive and
worst-scoring configuration). Token costs are stable across the three runs
(CV 0.9--17.6\%). In wall-clock time, all configurations finish well within
budget: mean session durations of 7.7--18.0 minutes consume only
6.4--15.0\% of the two-hour budget, and the medians (6.5--12.8 minutes)
confirm this is not driven by a few long outliers. Early termination, rather
than budget exhaustion, is the norm, echoing our finding that agents rarely
iterate with measurement (Appendix~\ref{app:case_study}).

Two caveats apply. First, billing weights are plan-specific: absolute costs
shift with pricing, but the scaffold ranking (Claude Code $\gg$ OpenHands $>$
OpenCode) is structural and independent of the exact weights. Second, these
figures cover the nine baseline configurations only; the profiling-tool
augmentation performs additional measurement cycles and is excluded from
this analysis.

\begin{table}[t]
\centering
\footnotesize
\setlength{\tabcolsep}{4.5pt}
\caption{Per-instance cost and duration by configuration (three-run
averages over 65 instances). Billing (M): millions of billing-equivalent
tokens per instance. Shares decompose the billing-equivalent total into
output, uncached input, and cache-read contributions. Budget: mean session
duration as a percentage of the two-hour budget. Score/1M: average score per
1M billing-equivalent tokens. Configurations are ordered by average score,
as in Table~1.}
\label{tab:cost}
\begin{tabular}{llrrrrrrrr}
\toprule
& & \multicolumn{4}{c}{Billing shares (\%)} & \multicolumn{2}{c}{Duration (min)} & \\
\cmidrule(lr){3-6}\cmidrule(lr){7-8}
Scaffold & Model & Billing (M) & Out. & Inp. & Cache & Avg. & Med. & Budget (\%) & Score/1M \\
\midrule
OpenHands   & GLM      & 0.74 & 17.2 & 17.7 & 65.1 & 16.9 & 12.8 & 14.1 & 57.1 \\
OpenCode    & GLM      & 0.60 &  7.6 & 19.8 & 72.5 & 16.7 & 12.8 & 13.9 & 67.0 \\
Claude Code & Qwen     & 1.48 & 16.5 & 24.2 & 59.3 &  8.5 &  7.9 &  7.1 & 26.6 \\
OpenHands   & Qwen     & 0.51 & 17.5 & 39.0 & 43.6 & 10.3 &  9.3 &  8.6 & 76.5 \\
Claude Code & GLM      & 0.75 & 16.5 & 31.5 & 52.0 &  7.7 &  6.5 &  6.4 & 50.9 \\
OpenCode    & Qwen     & 0.45 & 11.2 & 41.9 & 46.9 & 15.9 &  9.8 & 13.2 & 70.2 \\
OpenHands   & DeepSeek & 0.49 & 17.9 & 22.1 & 60.0 & 11.0 & 10.2 &  9.2 & 60.6 \\
OpenCode    & DeepSeek & 0.35 & 16.1 & 26.8 & 57.1 & 18.0 & 12.4 & 15.0 & 84.2 \\
Claude Code & DeepSeek & 1.69 & 18.3 & 32.4 & 49.3 &  8.5 &  7.9 &  7.1 & 13.9 \\
\bottomrule
\end{tabular}
\end{table}


\section{Measurement Protocol and Score Robustness}
\label{app:protocol}

\paragraph{Measurement protocol.}
For each instance, the reference latency $L_R$ and the degraded-baseline
latency $L_D$ were each measured three times before any agent run and
recorded as the average, after which they were frozen. The same recorded
pair is reused for all 27 configurations and all three rounds, so that
every one of the 1{,}755 trials is evaluated against an identical
numerical baseline; this freezing favors comparability across the full
experiment over per-batch re-measurement. Each submitted deployment is
evaluated by running the standard mixed workload three times and reporting the
mean of the three iterations.

\paragraph{Workload, gates, and resource budget.}
The evaluation workload issues requests at a fixed rate with a fixed
endpoint mix and input dataset, identical across all instances and
rounds; each measurement runs for the same duration per deployment.
Services run in isolated containers with fixed per-service CPU and
memory limits on a dedicated host, with no co-located load during
measurement. The error-rate gate fails any submission whose error rate
under load exceeds that of the injected (degraded) baseline by more
than one percentage point; timed-out requests are counted as errors and
excluded from the latency average. Agents may merge or split services
only when the aggregate CPU and memory budget of the system does not
exceed that of the original deployment.

\paragraph{Robustness of the success criterion.}
A positive score requires the measured optimized latency $L_O$ to fall
below the degraded baseline $L_D$, but a small positive difference could
in principle reflect measurement noise rather than a genuine improvement.
Two analyses suggest the reported results are not driven by such noise.
First, positive trials are far from the decision boundary: among 773
positive trials, only 1.2\% improve latency by less than 1\% and 6.3\% by
less than 5\%, while the median improvement is 62.8\%. Second, we
recomputed all statistics under success thresholds of $\delta \in
\{2\%, 5\%, 10\%\}$ minimum relative improvement: the positive rate
decreases modestly (44.0\% to 39.9\% at $\delta=10\%$), and all
model-level conclusions are unchanged (GLM and Qwen remain significantly
ahead of DeepSeek, $p<0.005$). Full recoveries are not knife-edge
either: among trials reaching $L_O \le L_R$, the median $L_O/L_R$ is
0.82, and 72.9\% land at or below $0.9\,L_R$.

\paragraph{Supplementary score statistics.}
To distinguish ``partially closing a large injected gap'' from
``approaching reference performance,'' we report the recovery score
together with the ratios $L_O/L_D$ and $L_O/L_R$. Overall, 19.8\%
of trials reach $L_O \le 1.1\,L_R$ and 24.6\% reach $L_O \le 1.2\,L_R$.
Among the 480 partial-score trials, 29.6\% already lie within
$1.2\,L_R$, while the remainder recover a substantial fraction of the
injected gap without reaching the reference; the median relative
improvement of partial trials is 51.5\%.

\end{document}